# Anisotropic terahertz optostriction in group-IV monochalcogenide compounds

Kun Liu, Jian Zhou[*]

*Center for Alloy Innovation and Design, State Key Laboratory for Mechanical Behavior of Materials, Xi'an Jiaotong University, Xi'an 710049, China*

## Abstract

Terahertz (THz) technology is a cutting-edge scheme with various promising applications, such as next generation telecommunication, non-destructive evaluation, security check, and in-depth characterization, owing to their sensitivity to material geometric change and good transparency. Even though tremendous progresses have been made during the past decade, exploration the mechanisms of THz-matter interaction microscopically is still in its infancy. In this work, we use thermodynamic theory to show how THz illumination deforms materials and use group-IV monochalcogenide compounds to illustrate it. According to our first-principles density functional theory calculations, THz light with intermediate intensity (~$10^9$ W/cm$^2$) could yield elastic deformations on the order of ~0.1%, depending on laser polarization direction. Large anisotropic opto-mechanical responses are also revealed. Finally, we show that such strain can be detected via measuring the layer-resolved shift current under a probe light irradiation.

---

[*] jianzhou@xjtu.edu.cn

**Introduction**

Mechanical deformation of a material is an efficient and facile approach to engineer its physical and chemical properties. For example, it has been widely studied and proved that elastic strain engineering is promising for electronic band structure manipulation, exciton formation and flow, topological phase transition, and controlling magnetic configurations [1-7], etc. Opto-mechanical approaches in materials, which converts photonic energy into mechanically elastic or plastic energy, is an unprecedented way to induce mechanical deformation through a noncontacting approach [8-10]. It could avoid traditional direct contacts between the tip and samples, which may introduce unwanted impurities and reduce the reversibility. In addition, optical irradiation approaches possess different degrees of freedom, such as polarization, frequency, incident angle, and intensity, which provide a versatile route to manipulating the material properties and applications. Thus, light illumination is advantageous for its easy tunability and noninvasive nature.

When the light frequency is comparable with the bandgap in a semiconductor, photon absorption occurs which excites electrons from the valence band to conduction band. The excited electron and hole could recombine via a non-radiative process, scattered by phonons, and produce photoconversion heat which enhances temperature. It may reduce the device performance and reversibility. Hence, low incident light frequency in the far-infrared range (on the order of a few terahertz, THz) becomes attractive since it is less susceptible to lattice damage. THz radiation is now a cutting-edge technology and has found various potential applications, such as information and telecommunication devices, non-destructive evaluation, food quality examination, environmental control, and ultrafast computing [11-13]. Over the past few years, THz impacts on the material geometric and electronic properties has been attracting great attention. In some cases, the THz light excites the infrared (IR) active phonon through resonant scattering, and nonlinear phonon interactions furthermore modulates the geometric lattice and electric polarization [14-17]. In addition, as the development of strong-field THz light source, off-resonant THz scattering [18-20] could also tune the ionic, electronic, and magnetic feature. The mechanical deformation under such off-resonant condition is still an open question and remains further investigation.

In this work, we show that the THz optics could induce structural elastic deformation, which is termed as THz optostriction (TOS) effect [schematically shown in Fig. 1(a)]. We use thermodynamic optostrictive theory that has been developed in the infrared frequency region [8] to show how TOS can be estimated, which requires computations of electronic and ionic susceptibility variations under stress. All of these functions can be directly computed according to first-principles density functional (perturbation) theory calculations. We then illustrated it in the group-IV monochalcogenide compounds (bulk GeS, GeSe, SnS, and SnSe, in their ground state *Pnma* phase), which are composed by antiferroelectrically stacked layers [21]. Their bulk structures are centrosymmetric with zero spontaneous electric polarization ($\vec{P}^s = 0$) [22]. They have been receiving great attention with quite a few promising properties and behaviors, such as high figure of merit thermoelectric effect, ultrafast phase transition, and good oxidation resistance [23-29]. Here, we perform first-principles calculations and suggest that all these structures show a large anisotropic TOS effect under linearly polarized THz light (LPTL) irradiation. We also explore their nonlinear optical responses by calculating the bulk photovoltaic effect under infrared-to-ultraviolet linearly polarized light (LPL), namely, shift current (SC) generation. Even though the *Pnma* is centrosymmetric, which forbids finite net SC in its bulk, we show that there are hidden layer-dependent SCs that flow oppositely in the neighboring layers. Such layer-dependent SCs can be effectively manipulated via TOS strains, yielding an all-optical approach to generate and control photocurrents in these systems.

**Theoretical and Computational Details**

*Thermodynamic theory of TOS effect.* According to the electromagnetics theory, there are four typical frequency regimes for an incident light with frequency $\omega$. The first one is dispersion regime where $\omega < \omega_t - \Delta\omega$ ($\omega_t$ refers to optical transition frequency, and $\Delta\omega$ is its broadening bandwidth), where optical dispersion dominates with marginal light absorption. This is because the electronic displacement is in phase with the electromagnetic wave while the amplitude of displacement is relatively small. The real part of dielectric function increases slightly as incident frequency increases, leading to light dispersion. The second one is absorption regime, $\omega_t - \Delta\omega < \omega < \omega_t + \Delta\omega$, where strong light absorption occurs and photoconversion heat would be produced. The reflection regime lies in the range of $\omega_t + \Delta\omega < \omega < \omega_p$ ($\omega_p$ is plasmon frequency). Here the absorption is again marginal but the real part of dielectric function is negative, leading to large reflection coefficient (assuming incidence from vacuum). The fourth one is transparent regime with $\omega > \omega_p$ where the optical transmission is dominant [30]. In the last two regimes, the light-matter interaction is very low, due to faster optical field oscillation than the typical characteristic time of materials. In the current situation, THz frequency is comparable to (or even lower than) typical phonon frequency, and is usually lower than electron interband transition frequency (note that group-IV monochalcogenides are semiconductors). We expect that THz optics could interact with both electron and ion subsystems in the group-IV monochalcogenides. Thus, we are now dealing with the low frequency dispersion regime if the infrared (IR) active phonon frequency is off-resonant with THz source. In this case, THz optics can be treated using alternating electric field, $\vec{\mathcal{E}}(t) = \text{Re}(\vec{E}e^{-i\omega_0 t})$, where $\omega_0$ is on the THz frequency order and is chosen to be 1 THz in the current study. For the bulk system where surface effect can be omitted, one uses electric field $\vec{\mathcal{E}}$ as the natural variable and evaluates the total Gibbs free energy (GFE) density variation under LPTL irradiation (Einstein summation convention is used) [31,32]

$$d\mathcal{G} = -S_{ij}dX_{ij} - \langle P_i^s d\mathcal{E}_i \rangle - \varepsilon_0 \varepsilon'_{ij}(\omega_0)\langle \mathcal{E}_i^* d\mathcal{E}_j \rangle \qquad (1)$$

Here $\varepsilon_0$ is the vacuum permittivity and $\langle * \rangle$ denotes the time average of quantity $*$. Isothermic condition is assumed, so that temperature effect is not considered here. The first term evaluates elastic contributions to GFE density, where $S_{ij}$ and $X_{ij}$ are strain and stress ($i, j$ = x, y, z) components, respectively. Positive (negative) strain represents tensile (compressive) deformation. The second term measures the interaction between the spontaneous polarization $\vec{P}^s$ and LPTL field, and the third term is the interaction between induced electric displacement and LPTL (note that $\vec{D}^{ind} = \varepsilon_0 \overleftrightarrow{\varepsilon}' \cdot \vec{\mathcal{E}}$ where $\overleftrightarrow{\varepsilon}'$ is the real part of dielectric tensor incorporating both ion and electron contributions). Since spontaneous polarization $\vec{P}^s$ is generally time-independent when the incident electric field strength is lower than its coercive field, its time average value is zero. Taking time average of sinusoidal electric field $\vec{\mathcal{E}}(t)$, the third term becomes $-\frac{1}{2}\varepsilon_0 \varepsilon'_{ij}(\omega_0) E_i dE_j$. Then, Eq. (1) implies that alternating electric field changes GFE density, so that mechanical responses under LPTL would occur. One can define a LPTL induced TOS strain, which is quadratic on the electric field strength [8]

$$S_{ij} = M_{ijkl}(\omega_0) E_k E_l \tag{2}$$

Here the coefficient $M_{ijkl}$ can be explicitly written as

$$M_{ijkl}(\omega_0) = \frac{\partial^2 S_{ij}}{\partial E_k \partial E_l} = -\frac{\partial^3 \mathcal{G}}{\partial X_{ij} \partial E_k \partial E_l} = \frac{1}{2}\frac{\partial^2 D_k}{\partial X_{ij} \partial E_l} = \frac{\varepsilon_0}{2}\frac{\partial \varepsilon'_{kl}(\omega_0)}{\partial X_{ij}} \tag{3}$$

$M_{ijkl}(\omega_0)$ is TOS coefficient at frequency $\omega_0$. Once again, this process requires no direct photon absorption in the material, neither through generating electron-hole pairs nor exciting IR-active phonons. Thus, no significant heat will be produced during TOS process, and the material destruction can be greatly eliminated.

*Electron and ion contributed dielectric functions.* As discussed previously, the dielectric function at THz frequency composes both electron and ion contributions, namely, $\overleftrightarrow{\varepsilon}(\omega_0) = \overleftrightarrow{\varepsilon}^{el}(\omega_0) + \overleftrightarrow{\varepsilon}^{ion}(\omega_0)$. This is different from our previous work that infrared light only interacts with electron subsystems [8]. Both of these dielectric functions can be separately evaluated according to linear response theory in the first-principles calculation framework. We apply independent particle picture in the random phase approximation (RPA) to compute the electron contribution [33]

$$\varepsilon_{ij}^{\text{el}}(\omega) = \delta_{ij} - \frac{e^2}{\varepsilon_0} \int_{BZ} \frac{d^3\boldsymbol{k}}{(2\pi)^3} \sum_{c,v} \frac{\langle u_{v,\boldsymbol{k}}|\nabla_{k_i}|u_{c,\boldsymbol{k}}\rangle \langle u_{c,\boldsymbol{k}}|\nabla_{k_j}|u_{v,\boldsymbol{k}}\rangle}{\hbar\left(\omega_{c,\boldsymbol{k}} - \omega_{v,\boldsymbol{k}} - \omega - \frac{\text{i}}{\tau^{\text{el}}}\right)} \quad (4)$$

where $|u_{n,\boldsymbol{k}}\rangle$ is the periodic part of wavefunction for band $n$ ($c$ and $v$ representing conduction and valence band, respectively) at $\boldsymbol{k}$. The electron relaxation lifetime $\tau^{\text{el}}$ is determined by the quality of sample, electron-electron and electron-phonon scattering, etc. Note that such lifetime should depend on different states, but a rigorous evaluation is not possible in the current theoretical framework, even for a perfect crystal sample. In this work, we take a universal value of $\tau^{\text{el}} = 0.04$ ps, which is usually adopted by theoretical works and conservable to experimental observations.

The ionic contributed dielectric function can be calculated according to the long wavelength limit (near $\Gamma$ point) phonon spectrum and vibration modes [34]

$$\varepsilon_{ij}^{\text{ion}}(\omega) = \frac{1}{V} \sum_m \frac{\mathcal{Z}_{m,i}^* \mathcal{Z}_{m,j}^*}{\omega_m^2 - \left(\omega + \frac{\text{i}}{\tau^{\text{ion}}}\right)^2} \quad (5)$$

where $\mathcal{Z}_{m,i}^* = \sum_{\kappa,i'} z_{\kappa,ii'}^* u_{m,\kappa i'}$ is the $i$-th Born effective charge of mode $m$, which quantifies the coupling between the optical phonons and electric fields, where $\kappa$ is ion index and $u$ is mass normalized displacement. $z_{\kappa,ii'}^*$ is Born effective charge tensor component of ion-$\kappa$, which measures the change of polarization along the $i$-component and the ionic displacement in the $i'$-direction. The phonon lifetime $\tau^{\text{ion}}$ mainly depends on anharmonicity of vibrations (high order phonon interactions) and environmental effects, which is much longer (on the order of picoseconds) than electron lifetime. We follow previous works and take a universal value of 8 ps, which is conservable in most cases [35,36]. Note that since we are focusing on the THz region, the space-charge polarization (~$10^5$ Hz) and dipolar polarization (~$10^9$ Hz) contributions can be safely omitted.

The real part of dielectric function, $\varepsilon'(\omega)$, denotes the optical dispersion effect. According to our previous analysis, it affects the GFE of the system. On the other hand, the imaginary part, $\varepsilon''(\omega)$, corresponds to light absorption. The phonon and electron lifetimes affect the peak height and broadening in the imaginary part of dielectric function. According to the Kramers-Kronig relation, the exact values of $\tau^{el}$ and $\tau^{ion}$ do not affect the real part value $\varepsilon'(\omega)$ when light absorption is negligibly small. We performed test calculations and find that the exact values of phonon and electron lifetimes does not change the values far below the resonant frequency (Supplemental Material), consistent with previous argument. Hence, we use the above mentioned universal lifetime values to conduct our calculations, as long as the frequency is chosen away from resonant absorption regime ($\varepsilon''(\omega_0) \sim 0$).

*Layer-resolved bulk photovoltaic effect.* Bulk photovoltaic effect under LPL (above bandgap), namely, SC density can be generated according to [37]

$$j_{sc}^a = 2\sigma_{bb}^a(0;\omega,-\omega)E^b(\omega)E^b(-\omega) \quad (6)$$

where $a$ and $b$ are Cartesian indices and $E(\omega)$ is the Fourier component of the optical electric field at angular frequency $\omega$ (above bandgap). According to the quadratic Kubo response theory, the coefficient $\sigma_{bb}^a$ can be written in various forms. Here we adopt the velocity gauge representation [38]

$$\sigma_{bb}^a(0;\omega,-\omega) = \frac{e^3}{2\omega^2\hbar^2}\int_{BZ}\frac{d^3\boldsymbol{k}}{(2\pi)^3}\text{Re}\sum_{m,n,l}\frac{f_{lm}v_{lm}^b}{\omega_{ml}-\omega+i\xi}\left[\frac{v_{mn}^a v_{nl}^b}{\omega_{mn}+i\xi} - \frac{v_{mn}^b v_{nl}^a}{\omega_{nl}+i\xi}\right] \quad (7)$$

where $f_{lm} = f_l(\boldsymbol{k}) - f_m(\boldsymbol{k})$ and $\hbar\omega_{lm} = \hbar\omega_l(\boldsymbol{k}) - \hbar\omega_m(\boldsymbol{k})$ are occupation and eigenenergy difference between band-$l$ and $m$ at $\boldsymbol{k}$, respectively. $v_{mn}^i = \langle m\boldsymbol{k}|\hat{v}^i|n\boldsymbol{k}\rangle$ is interband velocity matrix at $\boldsymbol{k}$. $\xi$ is phenomenological damping parameter incorporating disorder, impurity, and scattering effects, taking to be 0.02 eV in the current calculation. In centrosymmetric systems, such as *Pnma* structure in our case, $\sigma_{bb}^a(0;\omega,-\omega)$ is symmetrically forbidden with its value being exactly zero. However, one notes that each van der Waals (vdW) layer of group-IV monochalcogenides is non-centrosymmetric, hence we expect that layer-resolved SC exists, which could be furthermore modulated under TOS strains. Owing to the spatial separation of vdW gap, the SC in each layer can be individually measured experimentally. In order to evaluate SC response of each layer, we define a layer projection operator $\hat{P}_\ell = \sum_{i\in\ell}|\psi_i\rangle\langle\psi_i|$, where $|\psi_i\rangle$ is Wannier function centered in the $\ell$-th layer of the system ($\ell = 1, 2$).

Then we replace velocity operator $\hat{v}^a$ by $\hat{P}_\ell \hat{v}^a$, which corresponds to the current flowing in the $\ell$-th layer. In this way, a layer-resolved SC conductivity response function $\sigma_{bb}^a(0;\omega,-\omega)@\ell$ can be obtained.

*Density functional theory calculations.* The first-principles DFT calculations are performed in the Vienna *Ab initio* Simulation Package (VASP) [39], with the exchange-correlation potential treated by the generalized gradient approximation (GGA) in the Perdew–Burke–Ernzerhof (PBE) form [40]. The projector augmented wave (PAW) method and planewave basis set are applied to treat the core and valence electrons, respectively [41]. The kinetic cutoff energy of the planewave basis is set to be 350 eV. We use Monkhorst-Pack *k*-point sampling method with 9×9×3 grids to represent the first Brillouin zone [42]. The convergence threshold for the total energy and force components are set to be $10^{-9}$ eV and $10^{-7}$ eV/Å, respectively. The spin-orbit coupling (SOC) is self-consistently included in all calculations. We adopt the DFT-D3 scheme with Becke-Jonson damping dispersion correction to describe vdW interactions [43]. Phonon dispersion and Born effective charge of each ion are calculated using the finite displacement method within a (3×3×1) supercell [44]. Maximally localized Wannier functions are fitted in the Wannier90 code [45,46], to evaluate the SC response function.

**Results**

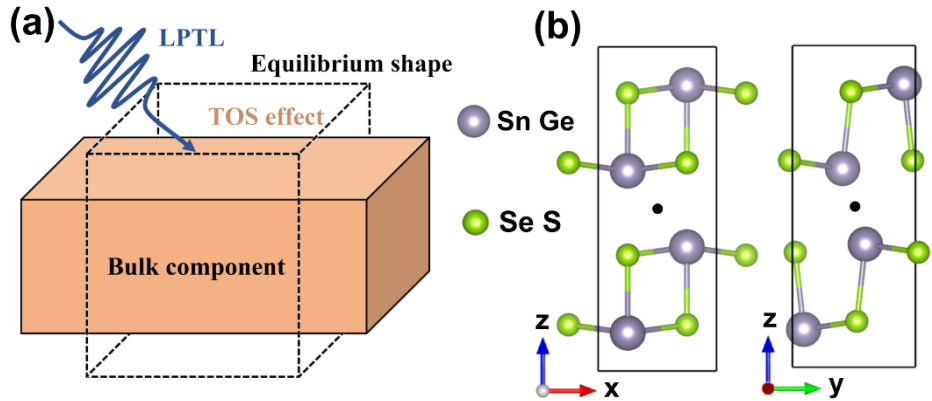

FIG. 1. (a) Schematic plot of LPTL induced TOS effect. (b) Atomic structure of *Pnma* group-IV monochalcogenide compounds viewed from [010] and [100] directions. Solid rectangles represent unit cell and the black circle dots denote the spatial inversion center.

*Mechanical and electronic properties.* The atomic structure of group-IV monochalcogenide compounds is shown in Fig. 1(b). The unit cell contains two layers, and each layer exhibits an in-plane ferroelectricity. According to structural optimizations, the layers are separated by a vdW gap of ~2.7 Å. The two vdW layers are stacked anti-ferroelectrically (interlayer), thus the total net polarization is zero and the structure possesses an inversion center in the vdW gap. We calculate the elastic constant components of group-IV monochalcogenides, which are tabulated in Table I. One sees clear mechanical anisotropy in these systems. Next, we compute their electronic band structure, as shown in Fig. 2. All these systems possess a bandgap of $0.5 - 1.2$ eV, which are much larger than the THz frequency range. Hence, according to Eq. (4), the imaginary part of $\varepsilon_{ij}^{\text{el}}(\omega_0)$ is zero (no direct interband transition), and the real part of $\varepsilon_{ij}^{\text{el}}(\omega_0)$ almost keeps a constant in the THz range ($\omega_0 \sim 1$ THz). We list their values in Table II. These results are consistent with previous works [47-50].

Table I. The mechanical elastic constants (in GPa) of different group-IV monochalcogenide compounds.

|  | $C_{11}$ | $C_{22}$ | $C_{33}$ | $C_{12} = C_{21}$ | $C_{13} = C_{31}$ | $C_{23} = C_{32}$ |
|---|---|---|---|---|---|---|
| GeS | 98.37 | 49.41 | 77.65 | 44.35 | 16.55 | 15.42 |
| GeSe | 90.39 | 46.07 | 79.72 | 42.39 | 15.27 | 16.55 |
| SnS | 82.23 | 44.85 | 80.55 | 36.54 | 19.92 | 23.59 |
| SnSe | 66.94 | 42.75 | 80.13 | 32.11 | 12.47 | 16.54 |

Table II. Electron contributed dielectric constant tensor ($\varepsilon^{\text{el}}$) components at the equilibrium structure of group-IV monochalcogenides. Note that the nondiagonal terms are all zero due to the *Pnma* symmetry constraints.

|  | $\varepsilon_{xx}^{\text{el}}$ | $\varepsilon_{yy}^{\text{el}}$ | $\varepsilon_{zz}^{\text{el}}$ |
|---|---|---|---|
| GeS | 11.90 | 11.13 | 11.78 |
| GeSe | 16.81 | 15.43 | 15.68 |
| SnS | 17.85 | 15.53 | 15.99 |
| SnSe | 24.08 | 22.07 | 20.72 |

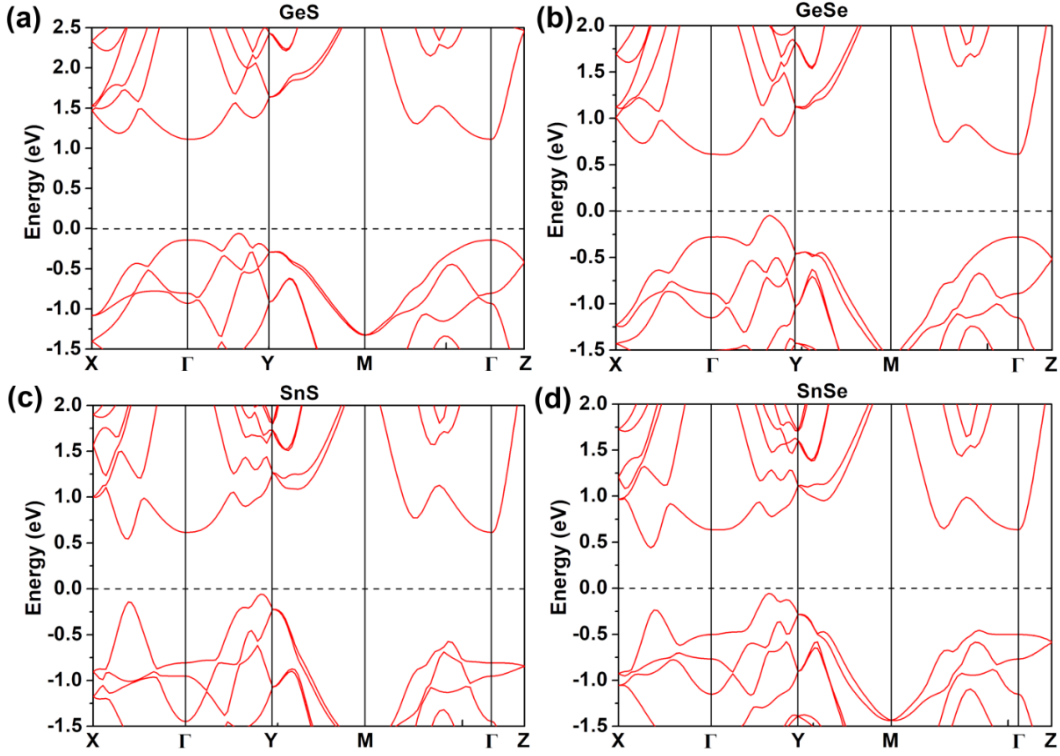

FIG. 2. Electron band dispersion of (a) GeS, (b) GeSe, (c) SnS, and (d) SnSe compounds along the high symmetric ***k***-path at their equilibrium state. The direct coordinates are $\Gamma = (0, 0, 0)$, $X = (0.5, 0, 0)$, $Y = (0, 0.5, 0)$, $M = (0.5, 0.5, 0)$, and $Z = (0, 0, 0.5)$.

*Phonon dispersion.* We then calculate their phonon dispersions, which are plotted in Fig. 3. One observes no imaginary modes throughout the Brillouin zone, consistent with the fact that *Pnma* phase is dynamically stable [black curves in Figs. 3(a)-(d)]. Compare with the lattice constants, the light wavelength is much longer (on the order of a few hundred μm). Hence, only long wavelength ($q \to 0$) modes contribute to optical response functions. The Γ-centered optical branch modes can be decomposed according to the irreducible representation as

$$\Gamma_{op}(D_{3h}) = 3B_{1u} \oplus B_{2u} \oplus 3B_{3u} \oplus 2B_{1g} \oplus 4B_{2g} \oplus 2B_{3g} \oplus 2A_u \oplus 4A_g \qquad (8)$$

The $B_{2u}$, $B_{1u}$, and $B_{3u}$ modes are IR-active, which generate electric polarizations along the *x*, *y*, and *z* directions, respectively. Hence, only these modes would contribute to the dielectric functions while all other modes are silent. We denote the dominate modes for each direction in the phonon dispersion plots, and their vibration displacements are schematically shown in Figs. 3(e)-(g). Since the group-IV monochalcogenides are ionic compounds, we add the non-analytic corrections to the eigenfrequencies [34,51]. One observes clear LO-TO splitting effects near Γ [red dashed curves in Figs. 3(a)-(d)], which agree well with previous results [52]. The formation of phonon polariton changes the LO frequency of IR-active modes, while the TO modes contribute to the LPTL response and there is marginal change of TO frequencies according to our calculations.

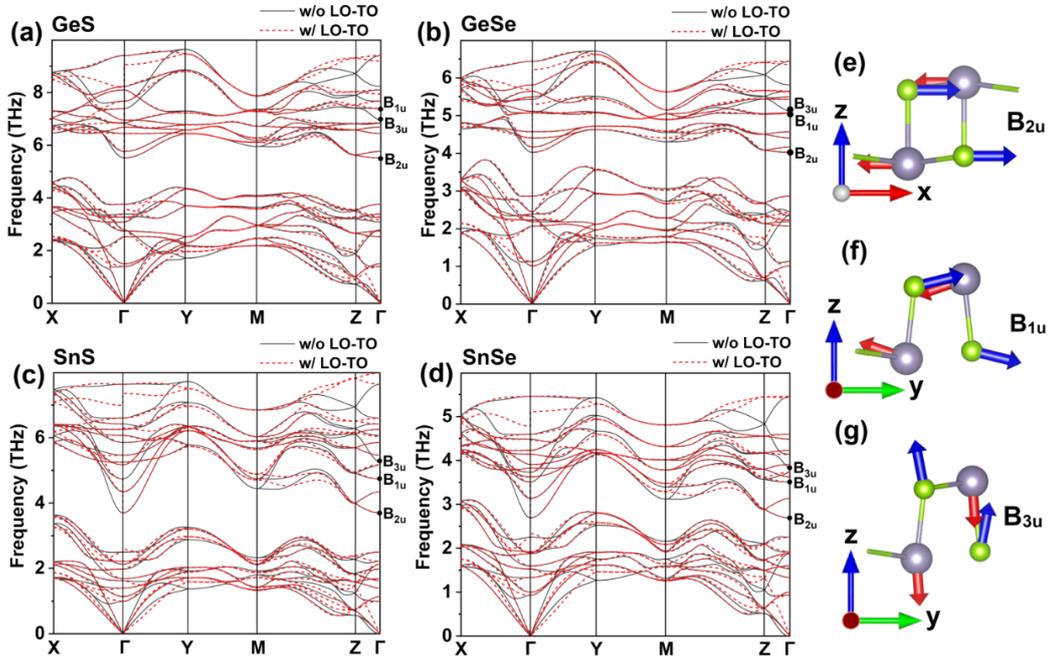

FIG. 3. Phonon dispersion of (a) GeS, (b) GeSe, (c) SnS, and (d) SnSe along the high symmetric ***k***-path. Red dashed and black solid curves represent the phonon dispersions with and without including non-analytic corrections, respectively. (e)-(g) show three strongest IR-active displacement modes along the *x* ($B_{2u}$), *y* ($B_{1u}$), and *z* ($B_{3u}$) axis without LO-TO splitting, respectively.

*Dielectric function variations under stress.* We then compute the ion contributed dielectric functions $\varepsilon_{ij}^{ion}(\omega)$, adding which to the electron contributions $\varepsilon_{ij}^{el}(\omega)$ yield total $\varepsilon_{ij}(\omega)$. Here only diagonal components are considered. Taking SnSe as an exemplary system, we plot its real and imaginary parts of $\varepsilon_{ii}(\omega)$ in Fig. 4. One can clearly see that at the equilibrium state, each $\varepsilon''$ component ($\varepsilon''_{xx}$, $\varepsilon''_{yy}$, and $\varepsilon''_{zz}$) has a dominant peak at 2.7, 3.5, and 3.9 THz, respectively. The corresponding real part of dielectric functions jump around these frequencies, arising from the Kramers-Kronig relation. Below all of these peaks, the real part of dielectric functions keeps constants. For example, at $\omega_0 = 1$ THz (or around this value), the $\varepsilon'_{xx}(\omega_0) = 85.4$, $\varepsilon'_{yy}(\omega_0) = 53.6$, $\varepsilon'_{zz}(\omega_0) = 41.8$, showing obvious optical anisotropy.

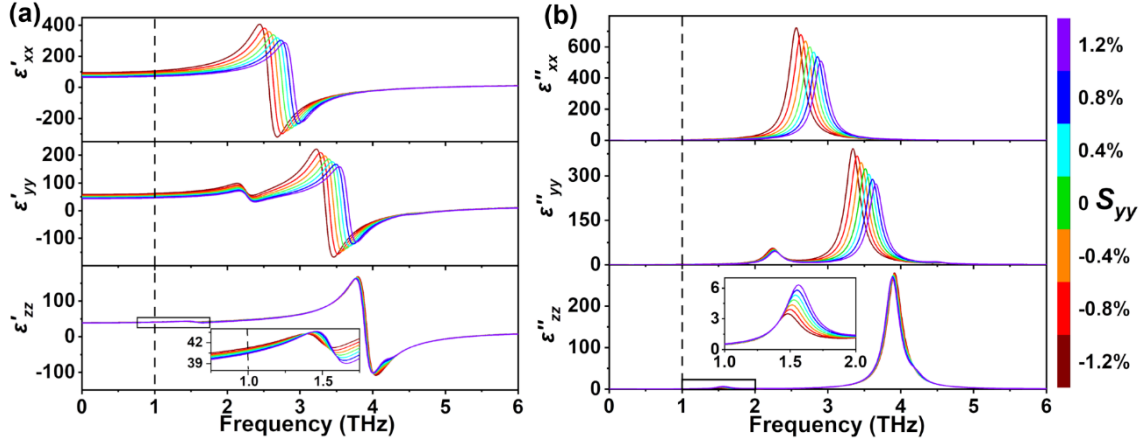

FIG. 4. (a) Real and (b) imaginary part of SnSe dielectric function components under uniaxial strain along $y$, $S_{yy}$. The color scheme of strain is shown in the colormap on the right. The inset in (a) and (b) are enlarged views of the $\varepsilon'_{zz}$ and $\varepsilon''_{zz}$ curves. Vertical dashed lines denote 1 THz frequency.

In order to evaluate TOS coefficient $M_{ijkl}(\omega_0)$, we artificially deform the structure along the $x$, $y$, and $z$ directions. Note that in the elastic deformation region, mechanical strain $S$ (as in Fig. 4) is proportional to stress $X$ [as used in Eq. (3)]. Both electron and ion contributed dielectric functions change under these deformations, owing to the strain-engineered electron and phonon dispersions. We use $S_{yy}$ strains to illustrate the results. The peak positions of $\varepsilon''_{xx}$ and $\varepsilon''_{yy}$ shift toward lower frequency and peak intensity becomes larger under compressive strain. These would yield larger real part values $\varepsilon'_{xx}$ and $\varepsilon'_{yy}$ in the lower frequency (e.g., around 1 THz). For $\varepsilon'_{zz}$, the increase is much smaller (under compression). Tensile mechanical strains reduce these real part values of dielectric functions. In order to show such tendency much clearer, we plot the $\varepsilon'_{ii}(\omega_0)$ at 1 THz variations as functions of uniaxial strains in Fig. 5. It is worth noting that these trends are also largely contributed by the electronic subsystem.

One clearly observes that in general, $\varepsilon'_{ii}(\omega_0)$ changes linearly with elastic strain.

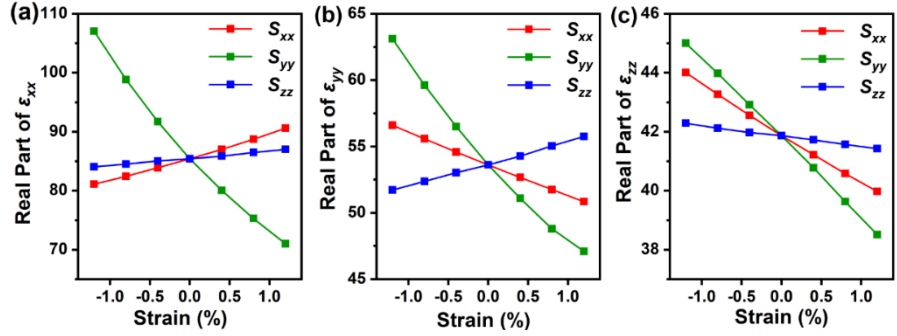

FIG. 5. Changes of $\varepsilon'_{ii}(\omega_0)$ ($i = x, y, z$) under intermediate elastic strains along the $x$, $y$, and $z$ directions for SnSe. Here $\omega_0$ is taken to be 1 THz. Note that around this frequency the results are the same, as the dielectric functions keep constants at low frequency.

*TOS coefficients.* For convenience, we use Voigt notation $M_{ij}$ to express TOS coefficient (for example, $M_{11} \leftrightarrow M_{xxxx}$ and $M_{12} \leftrightarrow M_{xxyy}$). We now apply Eq. (3) to calculate TOS coefficient components $M_{ij}(\omega_0)$. The calculated $M_{ij}(\omega_0)$ of all four group-IV monochalcogenides are listed in Table III. The positive (negative) values of $M_{ij}(\omega_0)$ indicates stretch (shrink) TOS stress along the $i$-direction when irradiating $j$-polarized LPTL (frequency $\omega_0 \sim 1$ THz). Remarkably, one sees that all the four systems show negative $M_{i3}(\omega_0)$ values ($i = 1, 2, 3$). This indicates that under $z$-polarized LPTL irradiating, the systems tend to shrink along all three directions. We expect one can precisely control the TOS strain by tuning the polarization and intensity of LPTL. This is due to the anisotropic optical responses and mechanical properties of the *Pnma* phase. Note that the mechanical Poisson's effect has not been taking into consideration yet, which will be added in the following discussion. Since the light adsorption at the 1 THz frequency is marginal, the penetration depth will approach infinite and the material is almost transparent to THz light. Hence, this TOS deformation can be universal in the system under irradiation, which is different from extrinsic deformations under external mechanical loads.

Table III. TOS coefficient $M_{ij}$ (in unit of Å$^2$/V$^2$) of group-IV monochalcogenides

under LPTL (at 1 THz frequency).

|  | x-LPTL | | | y-LPTL | | | z-LPTL | | |
|---|---|---|---|---|---|---|---|---|---|
|  | $M_{11}$ | $M_{21}$ | $M_{31}$ | $M_{12}$ | $M_{22}$ | $M_{32}$ | $M_{13}$ | $M_{23}$ | $M_{33}$ |
| GeS | 0.58 | −1.60 | −0.32 | −0.15 | −0.60 | −0.05 | −0.46 | −0.60 | −1.01 |
| GeSe | 0.75 | −3.06 | −0.36 | −0.26 | −1.20 | 0.05 | −0.56 | −0.64 | −1.07 |
| SnS | 1.31 | −7.73 | 0.27 | −0.81 | −3.45 | 0.49 | −0.73 | −0.27 | −2.07 |
| SnSe | 2.61 | −15.43 | 0.67 | −1.58 | −6.94 | 0.92 | −1.11 | −0.36 | −1.50 |

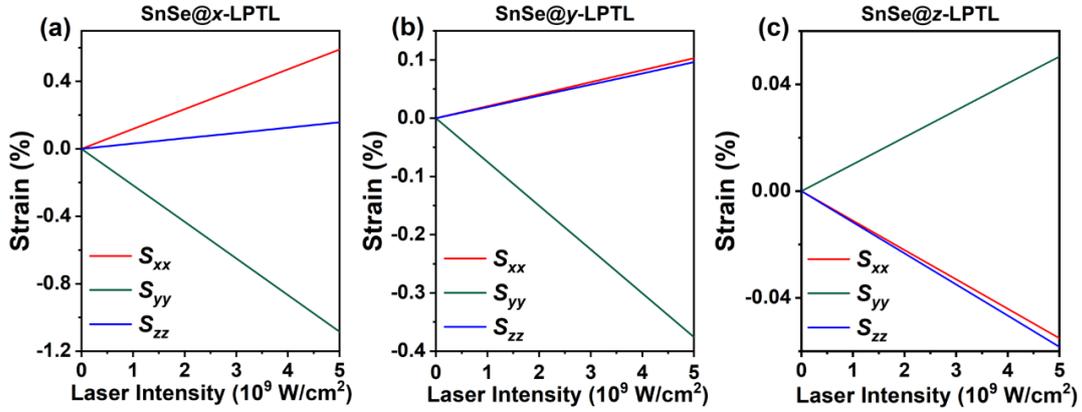

FIG. 6. Deformation strains of SnSe under both TOS and mechanical effects when irradiating under the (a) *x*, (b) *y*, and (c) *z*-polarized LPTL (at 1 THz frequency).

One has to note that the mechanical deformations are also strongly affected by its mechanical Poisson's effect. Considering this, the strains under LPTL becomes

$$S_{ii} = M_{ij}E_j^2 - \sum_{k \neq i}^{k} \nu_{ki} S_{kk} \qquad (9)$$

Note that the Poisson's ratio can be calculated according to the nondiagonal elastic constants, $\nu_{ij} = \frac{C_{ij}}{C_{jj}} (i \neq j)$. We set $S_{ii} = \mathfrak{M}_{ij}E_j^2$, where $\mathfrak{M}_{ij}$ serves as an effective THz opto-deformation response tensor, which includes mechanical Poisson's effect. The relationship between $\overleftrightarrow{\mathfrak{M}}$ and $\overleftrightarrow{M}$ is

$$\begin{bmatrix} \mathfrak{M}_{1j} \\ \mathfrak{M}_{2j} \\ \mathfrak{M}_{3j} \end{bmatrix} = \begin{bmatrix} 1 & \nu_{21} & \nu_{31} \\ \nu_{12} & 1 & \nu_{32} \\ \nu_{13} & \nu_{23} & 1 \end{bmatrix}^{-1} \begin{bmatrix} M_{1j} \\ M_{2j} \\ M_{3j} \end{bmatrix} \qquad (10)$$

Detailed derivations are shown in Supplemental Material. Taken SnSe as an example, we calculate the TOS strains under different laser intensity LPTL at 1 THz (Fig. 6). For example, the TOS strains along $x$ and $y$-axis are could reach 0.6% and −1.1%, respectively, under an $x$-polarized THz laser with intensity of $5 \times 10^9$ W/cm$^2$. This intermediate intensity can be well achieved in experiments which would not induce plastic structural phase transformation according to our estimation (Supplemental Material). The TOS strains of group-IV monochalcogenides (under laser intensity of $5 \times 10^9$ W/cm$^2$) are listed in Table IV. One sees that under $x$-LPTL illumination, the strain along the $y$-axis is the largest. Under the same light intensity, SnSe has the largest TOS induced deformation strain, while GeS has the smallest, and the strains of all four compounds can reach the order of $10^{-3}$. Compare with previous photostriction experiments (short wavelength light) for Si, Ge and SbSI (strain is below $10^{-4}$) [53-55], PbTiO$_3$ and BiFeO$_3$ perovskites (greater than $10^{-3}$ strain with much higher laser intensity, which in the order of $10^{11}$ W/cm$^2$) [56,57], this TOS induced strains can be orders of magnitude larger under the same laser intensity. In addition, the previously reported photostriction process highly depends on the excited carrier concentration [54,58,59], which may be reduced by the finite penetration depth and excitation limit of the host semiconductors.

Table IV. The TOS strain (under 1 THz light with its intensity of $I = 5 \times 10^9$ W/cm$^2$) of group-IV monochalcogenide compounds (unit of each strain value is $1 \times 10^{-3}$).

|  | $x$-LPTL | | | $y$-LPTL | | | $z$-LPTL | | |
| --- | --- | --- | --- | --- | --- | --- | --- | --- | --- |
|  | $S_{xx}$ | $S_{yy}$ | $S_{zz}$ | $S_{xx}$ | $S_{yy}$ | $S_{zz}$ | $S_{xx}$ | $S_{yy}$ | $S_{zz}$ |
| GeS | 0.82 | −1.33 | −0.03 | 0.08 | −0.30 | 0.02 | −0.10 | -0.02 | −0.35 |
| GeSe | 1.45 | −2.52 | 0.11 | 0.20 | −0.68 | 0.12 | −0.17 | 0.05 | −0.38 |
| SnS | 2.78 | −5.76 | 1.10 | 0.42 | −1.99 | 0.66 | −0.35 | 0.65 | −0.88 |
| SnSe | 5.89 | −10.84 | 1.57 | 1.03 | −3.76 | 0.96 | −0.55 | 0.50 | −0.58 |

*Layer-resolved SC generation.* The observation and detection of the TOS strain is not straightforward in practice. This *Pnma* phase is centrosymmetric, which does not show electrical polarization responses as in the piezoelectrics. Here we propose another optical approach, generating SC under LPL, to probe the deformations. According to nonlinear optical theory, SC produces in a single material under LPL illumination. The incident LPL photon energy should be larger than the direct bandgap of the semiconductor, thus we will consider IR to ultraviolet region. The photo-excited electron and hole may reside at different spatial positions (wavefunction center mismatch between the valence and conduction bands), then one can detect electric current. This is due to anharmonic motion of carriers, so that a general rule to realize SC is that the system lacks inversion symmetry. In the current case, even though the bulk *Pnma* is centrosymmetric, each vdW layer is ferroelectric which could conceive SC [60,61]. Hence, we calculate the SC conductivities $\sigma_{bb}^{y}(0;\omega,-\omega)$ ($b = x$ or $y$, LPL polarization) for SnSe. The *y*-direction is along the armchair buckling, parallel to in-plane electric polarization. Note that since each layer (and the bulk) contains a mirror symmetry $\mathcal{M}_x$, there will be no SC along *x*. We calculated the layer-resolved $\sigma_{bb}^{y}$ of the two vdW layers in the unit cell, as shown in Fig. 7(a). We can find that the $\sigma_{bb}^{y}$ in the two adjacent layers flow oppositely with same magnitude, yielding zero net SC in the whole compound. The layer-dependent SC can be effectively modulated via deformations. Taking uniaxial $S_{yy}$ strain as an example, the SC conductivity in the layer-1 variations (incident photon energies of 2.0 eV) are shown in Fig. 7(b). Remarkably, one can see that $\sigma_{xx}^{y}$ flips its flowing direction under tensile and compressive strains. Hence, this layer-dependent SC could be used to as a sensitive probe and characterize scheme for small deformations in these centrosymmetric materials [Fig. 7(c)]. Owing to the vdW gap between each layer, one could detect and measure the voltage of a single layer (or odd numbered layers in nanometer scale, but not the whole bulk sample). This can be done by carefully depositing electrodes on the surface layer of the system. It requires subtle experimental fabrication and rational design, which could be realized owing to the advances of the state-of-the-art nano- and micro-electronic techniques but is beyond the scope of our current work.

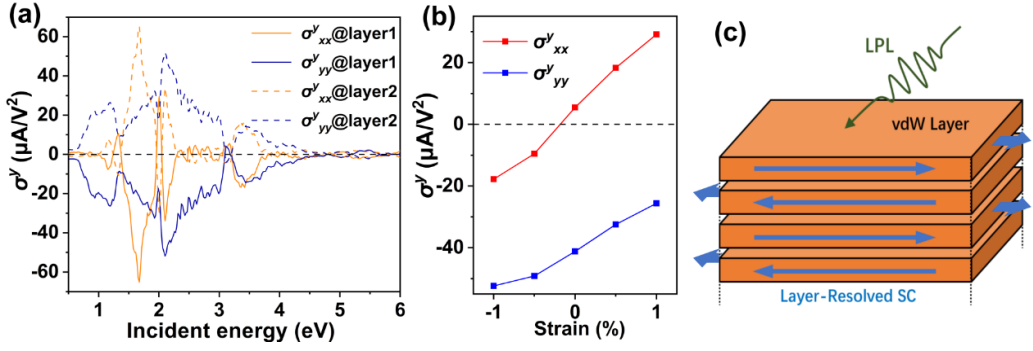

FIG. 7. (a) Layer-resolved SC conductivities under linearly polarized light of SnSe at its equilibrium structure. (b) Variation of $\sigma^y_{xx}$ and $\sigma^y_{yy}$ in the layer-1 as functions of strain $S_{yy}$ (from $-1\%$ to $1\%$). Incident photon energy of 2.0 eV is adopted. The currents in the layer-2 flow oppositely. (c) Schematic plot of SC in each vdW layer. The current flows along the armchair direction.

**Discussion**

According to the equipartition theorem, the thermal vibration energy of group-IV monochalcogenide compounds is on the order of $10^2$ J/cm³ at room temperature. While the energy scale of LPTL response in the current study is $10^1$–$10^2$ J/cm³, the TOS might be strongly affected by thermal fluctuation. The experimental measurement of TOS effect in group-IV monochalcogenides may be performed under low temperature. As for the detection resolution, current experimental technique can well-distinguish strains greater than 0.01% [62]. Nonetheless, the optical responses under THz light is affected by the frequency of IR-active modes, even though under off-resonant process in the current work. One could expect large TOS effects in other materials that has stronger THz responses and smaller mechanical elastic coefficients, and the measurement can be carried out under room temperature.

One may wonder if such mechanism can be used to calculate the TOS effect for metals or isolated molecules. Due to the presence of free electrons in metals, one may expect large absorption of THz light in metallic systems. According to Drude model, the complex form of electron conductivity is $\sigma(\omega) = \frac{ne^2\tau/m}{1-i\omega\tau}$, where $n$, $\tau$, and $m$ are carrier density, lifetime, and mass, respectively. The dielectric function is $\varepsilon(\omega) = \varepsilon_0 +$ $\frac{i\sigma(\omega)}{\omega}$. Hence, the real part of dielectric function is proportional to carrier density, and its lifetime dependence is complicated. In addition, note that the Drude model is essentially an intraband transition, and the interband transition also contributes to the conductance (and dielectric function). According to the linear response theory, the interband transition at low frequency tends to diverge for metals, which strongly depends on the carrier lifetime and the exact band character. Hence, an accurate evaluation of dielectric function is very challenging. In addition, electron-hole excitation occurs under THz field for metals, which adds another contribution to the Gibbs free energy. Due to strong light absorption, the penetration depth of THz in metals limits its effect in the bulk. Therefore, the metal responses would be very different from the semiconductors.

As for molecules, previous studies have reported that the properties of molecular chemical bonds can be changed under light, which would change the chemical reactions [63,64]. Similar as semiconductors or insulators, molecular systems possess a finite bandgap. Hence, we expect that similar approach can be applied to molecules. The difference between bulk and molecular systems lies at their boundary conditions. Under (alternating or static) electric field, the former one actually corresponds to closed circuit boundary condition which uses electric field as natural variable for thermodynamic analysis, while the latter one requires the electric displacement as the natural variable [32]. Therefore, the feasibility and further application of this optomechanical scheme remains to be studied in the future.

**Conclusion**

In conclusion, we predict that THz light could yield significant and anisotropic elastic deformations in group-IV monochalcogenide compounds. Based on thermodynamic theory analysis, we calculate the TOS effect by performing first-principles calculations. Since THz could penetrate inside the material without strong absorption, this process is non-destructive and will not cause overheating problems. Under intermediate light intensity, one could yield elastic strains on the order of 0.1%, which is observable experimentally and large enough for sensing and evaluations in practice. Finally, we propose that strains can be detected by measuring layer-resolved SC responses. The layer-resolved SC magnitude and direction can be effectively tuned by elastic strains. Our work provides an all-optical approach to generate and probe elastic strains using non-destructive THz technique and non-contacting nonlinear optical process.

**Acknowledgments.** This work was supported by the National Natural Science Foundation of China under Grant Nos. 21903063 and 11974270. The computational resources from HPC platform of Xi'an Jiaotong University are also acknowledged.